\documentclass[preprint,12pt]{elsarticle}

\usepackage{amssymb}
\usepackage{hyperref}

\journal{NIM A}

\begin{document}

\begin{frontmatter}



%\title{Title of Your Manuscript}
\title{A Liquid Xenon Positron Target Concept}


\author[inst1]{Max Varverakis}
\author[inst1]{Robert Holtzapple}

\affiliation[inst1]{organization={Department of Physics},%Department and Organization
            addressline={California Polytechnic State University}, 
            city={San Luis Obispo},
            postcode={93407}, 
            state={CA},
            country={USA}}

\author[inst2]{Hiroki Fujii}

\affiliation[inst2]{organization={Nishina Center},%Department and Organization
            addressline={RIKEN, 2-1 Hirosawa}, 
            city={Wako},
            postcode={351-0198}, 
            state={Saitama},
            country={Japan}}

\author[inst3]{Spencer Gessner\corref{cor1}}

\affiliation[inst3]{organization={SLAC National Accelerator Laboratory},%Department and Organization
            addressline={2575 Sand Hill Road}, 
            city={Menlo Park},
            postcode={94025}, 
            state={CA},
            country={USA}}

\cortext[cor1]{Corresponding author: sgess@slac.stanford.edu}

\begin{abstract}
Positron targets are a critical component of future Linear Colliders.
Traditional targets are composed of high-Z metals that become brittle over time due to constant bombardment by high-power electron beams.
We explore the possibility of a liquid xenon target which is continuosly refreshed and therefore not susceptible to the damage mechanisms of traditional solid targets. 
Using the GEANT4 simulation code, we examine the performance of the liquid xenon target and show that the positron yield is comparable to solid targets when normalized by radiation length.
Additionally, we observe that the peak energy deposition density (PEDD) threshold for liquid xenon is higher than for commonly employed metal targets, which makes it an attractive, non-toxic positron target alternative.
We develop parameter sets for demonstration applications at FACET-II and future Linear Colliders.
\end{abstract}



% \begin{keyword}
% %% keywords here, in the form: keyword \sep keyword
% keyword one \sep keyword two
% %% PACS codes here, in the form: \PACS code \sep code
% %% MSC codes here, in the form: \MSC code \sep code
% %% or \MSC[2008] code \sep code (2000 is the default)
% \end{keyword}

\end{frontmatter}


\section{Introduction}
Future linear colliders require approximately $10^{14}$ $e^+$ per second at the IP in order to achieve luminosities in excess of $10^{34}$ cm$^{-2}$s$^{-1}$~\cite{Seimiya2015}. Traditionally, positrons are produced by directing high energy electrons into a high-Z solid target, where positrons are created from the resulting electromagnetic shower.  Generating $10^{14}$ $e^+$ per second requires extremely high-power electron beams on target, which results in degradation of the targets over time~\cite{Bharadwaj2001}. Research into advanced positron sources has been recognized as an area-of-need for future accelerator research and development~\cite{Chaikovska2022,snowmassPosi}.

Previous experiments have explored alternatives to high-Z solid targets. Liquid mercury targets are commonly employed at neutron spallation sources~\cite{Kaminskas2018,Guan2018} and have also been investigated in the context of the International Linear Collider (ILC)~\cite{Mikhailichenko2006} and Muon Collider (MC)~\cite{McDonald:2009zz}. However, the vacuum requirements for the Hg vessel are more stringent for the ILC compared to spallation sources, and the toxicity of Hg presents unique hazards. Similarly, liquid lead targets have been considered for the Next Linear Collider (NLC)~\cite{Sheppard2002}, but were abandoned for reasons of toxicity. Low-Z targets have been considered for generating neutrons from liquid lithium~\cite{Feinberg2011}, generating muons from gaseous deuterium~\cite{Okita2020}, and generating muons from beryllium targets using a positron beam driver~\cite{Alesini2019}.

Xenon is a nonreactive, high-Z substance that can be converted into a dense liquid at relatively high temperatures. Liquid xenon (LXe) has several properties that make it attractive as a positron target. First, due to its density and high atomic number, LXe has a relatively short radiation length for a nonmetal. Second, LXe has a large heat of vaporization; it can absorb significant energy before vaporizing. The amount of energy that a target material can absorb is characterized by the peak energy deposition density (PEDD) and is usually capped at 35 Jg$^{-1}$ for metal targets~\cite{Ecklund1981}, but the PEDD can exceed 90 Jg$^{-1}$ in the case of LXe. Third, using LXe as a positron target removes the concern of long term degradation observed with solid positron targets because the LXe is continuously refreshed. Finally, LXe is non-toxic and requires much less safety infrastructure than liquid metal targets.

In this paper, we explore the use of a LXe positron target with the necessary flow rate to adequately handle the power deposited by a high-power electron beam. Previous studies at FACET-II~\cite{Yakimenko2019} considered a solid tantalum (Ta) target for in-situ positron generation for plasma wakefield acceleration experiments~\cite{Fujii2019}. We examine the viability of the LXe target by first comparing it with the Ta target and demonstrating that both targets achieve similar yield when normalizing by radiation length. We then examine the energy deposition in the LXe target and the windows of the encapsulating chamber. Finally, we analyze the performance of the LXe target using beam parameters from future Linear Collider concepts. 

\begin{figure}[htb]
    \includegraphics[width = \linewidth]{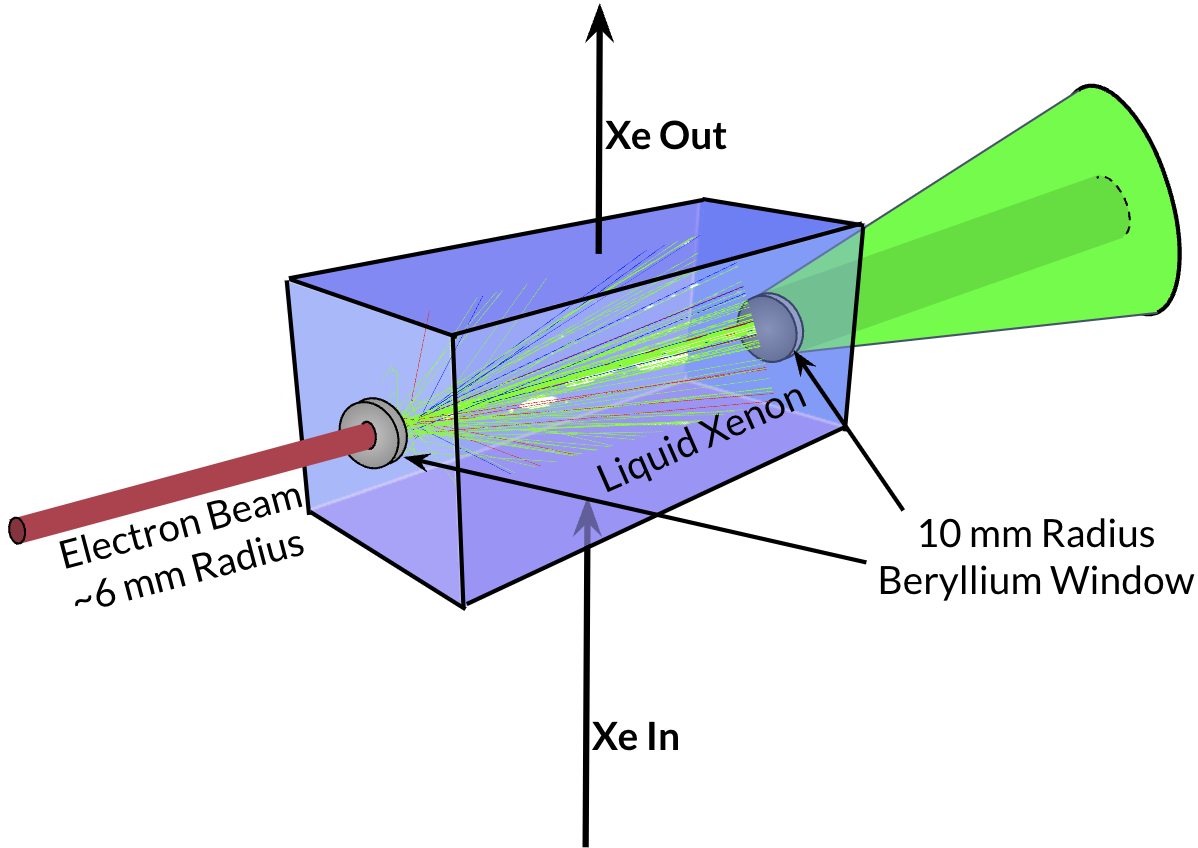}
    \caption{\label{fig:schem}The liquid xenon target setup. A 10 GeV electron beam is incident on the target from the left. The beam enters and exits the LXe target through a 10 mm diameter, 0.5 cm-thick beryllium window. The positrons exit the target and travel to the right.
    Fresh LXe is pumped vertically through the target chamber to replace the heated LXe.}
\end{figure}

\section{Comparisons of LXe and Ta Targets in GEANT4}
\label{sec:comp}

Figure~\ref{fig:schem} illustrates the LXe beam-target interaction simulated in  GEANT4~\cite{Geant4}. We use GEANT4 to quantify positron production yield and outgoing beam parameters from the electromagnetic shower for both the LXe and Ta targets. A previous study on positron production from Ta targets at FACET-II serves as a starting point for comparisons with the LXe target~\cite{Fujii2019}. We note that Ta targets have a similar radiation length compared to conventional tungsten-rhenium (W$_{75}$Re$_{25}$) targets.

For each run of the simulation, we sent $N_0$ = 10,000 macro particles onto the target with an incoming spot size of 6 mm. We investigated three different energies of the incident electron beam (3 GeV~\cite{Nagoshi2020}, 6 GeV~\cite{Seimiya2015,Tang1995}, and 10 GeV~\cite{Fujii2019}) and we scanned over the radiation length of the target from 0.5 radiation lengths to 8 radiation lengths. Table~\ref{tab:G4Params} shows the target material parameters used in the GEANT4 simulations.

\begin{table}[hb]
    \centering
    \begin{tabular}{cccc}
        \hline\hline
        \textbf{Material} & \textbf{Z} & \textbf{Density} [$\textrm{g} \cdot \textrm{cm}^{-3} $] & \textbf{Radiation Length} [cm] \\
        \hline
        Ta & 73 & 16.654 & 0.4094 \\
        % \hline
        LXe & 54 & 2.953 & 2.872 \\
        \hline\hline
    \end{tabular}
    \caption{\label{tab:G4Params}Parameters used in GEANT4 simulation when comparing targets. Material data is provided by the PDG~\cite{Workman:2022ynf}.}
\end{table}

\begin{figure}[htb]
    \includegraphics[width = \linewidth]{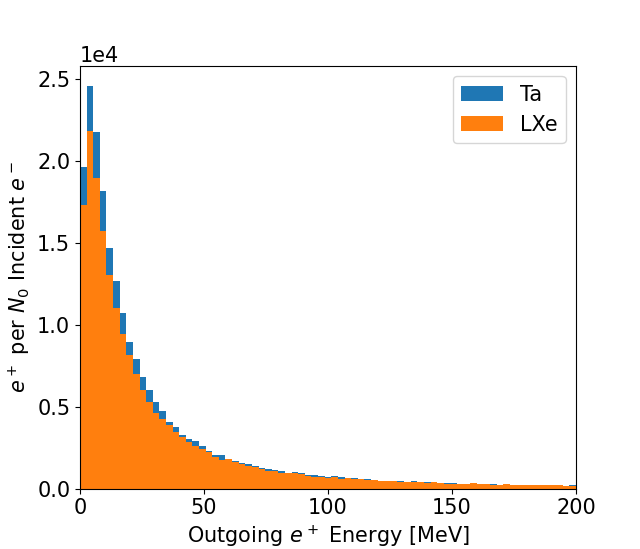}
    \caption{\label{fig:ESpectra}Energy spectrum of positrons generated in 5.5 radiation length LXe and Ta targets by a 10 GeV electron beam. Histogram counts are per $N_0$ = 10,000 incident $e^-$ per 2.67 MeV bin.}
\end{figure}

\begin{figure}[htb]
    \includegraphics[width = \linewidth]{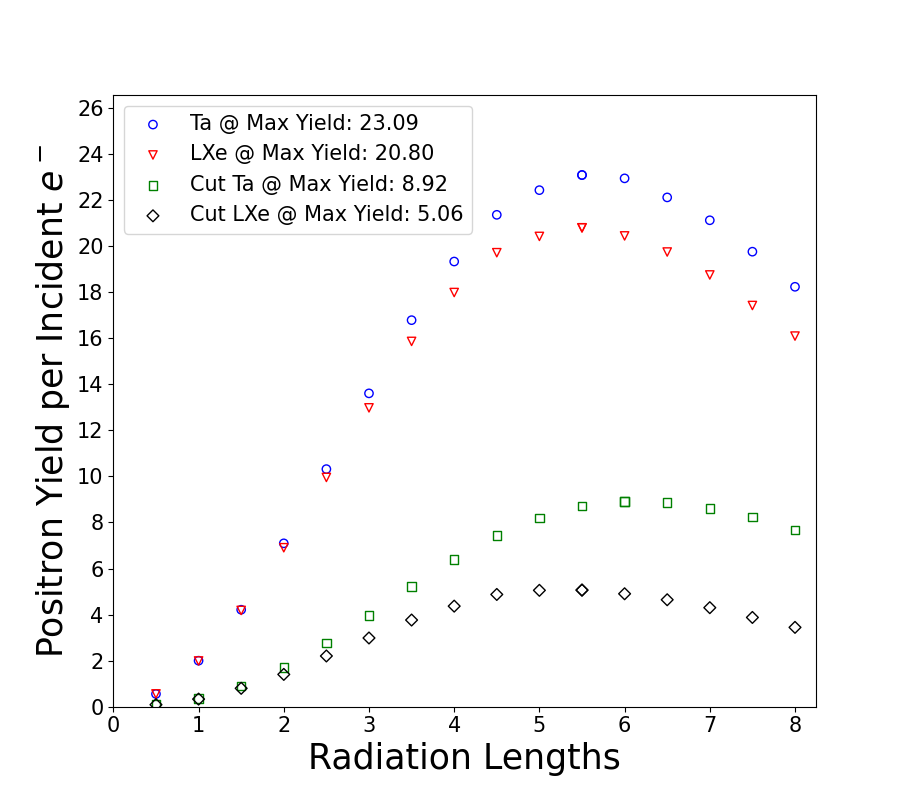}
    \caption{\label{fig:Yield} Positron yield per incident electron versus radiation length for LXe and Ta targets before and after particle selection is applied to outgoing positrons. The incident electron beam energy is 10 GeV. The cuts applied to the outgoing positrons require that their energy is greater than 2 MeV and less than 22 MeV, with a transverse offset less than 10 mm.}
\end{figure}

Figure~\ref{fig:ESpectra} compares the energy spectra of the outgoing positrons from the LXe and Ta targets at 5.5 radiation lengths. The spectra are nearly identical, with some loss of low-energy particles in the case of LXe due to the longer radiation length of the material.

The figure-of-merit used to compare the Ta and LXe targets is positron yield. We do not model the capture and acceleration sections after the target. Instead, we apply cuts on the phase space of the outgoing beam particles to simulate losses in particle capture and acceleration sections. Our energy acceptance window is from 2 MeV to 22 MeV. The 20 MeV window is consistent with previous positron source design work~\cite{Tang1995, Bayar2017, Nagoshi2020}. We assume a 10 mm radius cut on particles exiting the target, consistent with the aperture of the ILC flux concentrator~\cite{Nagoshi2020}.

 In addition to looking at the energy spectra of the outgoing positrons for each target material, Figure~\ref{fig:Yield} displays the positron yield as a function of radiation length for Ta and LXe. Without applying the particle selection cuts, our simulations show that the maximum yield for both Ta and LXe occur at a target depth of 5.5 radiation lengths and the overall yield agrees at the 10\% level. Application of the selection criteria reduces the yields for both Ta and LXe, but the reduction in yield is more significant for the LXe target because of the transverse cutoff. Once again, this is because the radiation length of LXe is 7 times that of Ta. For showers with similar angular divergence, the transverse width of the positron shower in LXe will be larger than in Ta.

Despite the larger shower width of LXe compared to Ta, the emittances for the outgoing positron beams are similar after selecting for energy and transverse offset. We use the following definition of normalized emittance:
\begin{equation}
     \varepsilon_{n,rms} = \frac{1}{m_0c}\sqrt[]{\left\langle x^2 \right\rangle \left\langle p_x^2 \right\rangle - \left\langle xp_x \right\rangle}. \label{eq:Emittance}
\end{equation}
At maximum positron yield (5.5 radiation lengths) for a 10 GeV input beam, the normalized transverse emittance for Ta is $38\textrm{ mm}\cdot\textrm{rad}$ while the transverse emittance for LXe is $36.5\textrm{ mm}\cdot\textrm{rad}$. The emittance values for Ta and LXe are nearly the same as a result of the particle selection cuts. The emittance is within the range of 9-60 mm$\cdot$rad that has been considered for the previous NLC, ILC, and CLIC designs~\cite{Tang1995, Seimiya2015, Vivoli2010}.

We note that the maximum yield for the LXe target after applying the particle selection cuts is 5, but the yield for the ILC and CLIC positron sources is typically closer to 1 when modeling the complete target-to-damping ring lattice~\cite{Nagoshi2020,Kamitani:492189}. The particle selection criteria we apply does not provide a complete accounting of losses in the transport, but we expect that our yield values provide sufficient overhead such that the final target-to-damping ring yield is greater than 1.

\section{Energy Deposition and Flow Rate Considerations for LXe}

\begin{figure}[htb]
    \includegraphics[width = \linewidth]{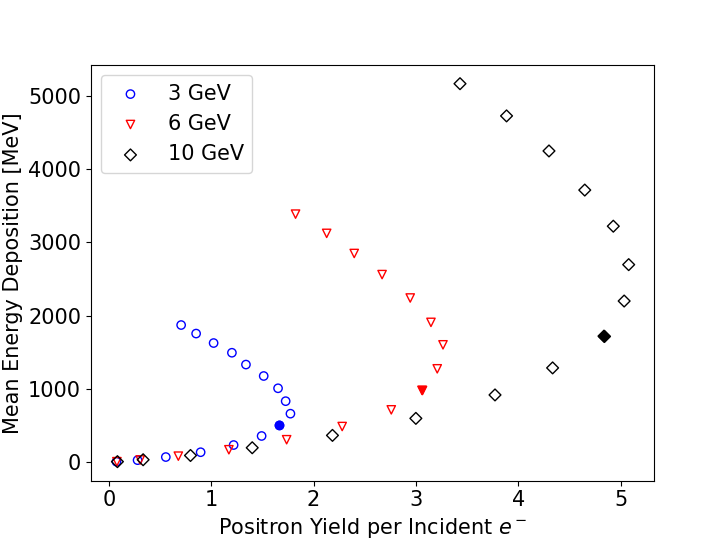}
    \caption{\label{fig:EDep}Mean energy deposition in LXe target per incident electron as a function of positron yield at three different beam energies. The points represent different radiation lengths for the LXe target. The filled-in markers indicate the operating radiation length selected for each beam energy, which is 3.5, 4.0, and 4.5 radiation lengths for a 3 GeV, 6 GeV, and 10 GeV beam, respectively.}
\end{figure}

In Section~\ref{sec:comp}, we used our GEANT4 simulations to demonstrate that the LXe target provides positron yields that are comparable to traditional targets. In order to demonstrate the advantages of the LXe target, we use the output of the GEANT4 simulations to analyze energy deposition in the LXe and the beryllium windows of the containment vessel. We use the energy deposited in the LXe target as a starting point for the LXe flow rate calculations. Figure~\ref{fig:EDep} displays the energy deposited in the LXe target per incident electron.

\begin{table*}
    \centering
        \begin{tabular}{ccccc}
            \hline\hline
            \textbf{Parameter} & \textbf{Unit} & \textbf{FACET-II}~\cite{FACET2016} & \textbf{ILC}~\cite{Nagoshi2020} & \textbf{C$^3$}~\cite{Bai2021} \\
            \hline
            Energy & GeV & 10 & 3 & 3 \\
            $e^-$/bunch & $10^{10}$ & 1.25 & 2.5 & 0.78 \\
            Bunches/train & & 1 & 1312 & 133 \\
            Rep. rate & Hz & 10 & 5 & 120 \\ \\
            \textbf{LXe Target} & & & & \\
            \hline
            Radiation Length & & 4.5 & 3.5 & 3.5 \\
            PEDD/train & $\textrm{J}\cdot\textrm{g}^{-1}$ & 0.080 & 78.8 & 2.49\\
            Flow rate & $\textrm{cm}^3\cdot\textrm{s}^{-1}$ & 0.12 & 57 & 36 \\ \\
            % \hline
            \textbf{Beryllium Window} & & & & \\
            \hline
                $E_{\textrm{dep}}$/train & J &  0.017 & 15.0 & 0.475 \\
                PEDD/train & J$\cdot \textrm{g}^{-1}$ & 0.058 & 52 & 1.6 \\
                $\Delta$T/train & K & 0.032 & 28.4 & 0.900 \\
                $\Delta$T/time & K$\cdot \textrm{s}^{-1}$ & 0.320 & 142 & 108 \\
            \hline\hline
        \end{tabular}
    \caption{\label{tab:BeamInfo}Electron beam parameters and associated LXe target quantities for FACET-II, ILC, and C$^3$ at the operating radiation lengths identified in Section~\ref{sub:Flow}.
    The beryllium window quantities are specific to a 10mm radius, 0.5 mm-thick Be disk.  Note that the window parameters correspond to the exit window only since the entrance window receives a significantly smaller energy deposit per incident electron.}
\end{table*}

\subsection{Calculating the LXe Flow Rate} \label{sub:Flow}

The LXe flow rate is determined by the amount of energy that a volume of LXe can absorb before evaporating. The heat of vaporization of LXe is $\Delta \textrm{H}_{\textrm{vol}} = 284.2$ J$\cdot \textrm{cm}^{-3}$, or $\Delta \textrm{H}_{\textrm{mass}} = 96.2$ J$\cdot \textrm{g}^{-1}$. We define $\textrm{PEDD}_{\textrm{max}} = \Delta \textrm{H}_{\textrm{mass}}$, the maximum value of energy that can be deposited in the LXe target by a train of bunches before the target volume must be refreshed.

The volume of LXe that the incoming beam interacts with depends on the thickness of the LXe target. In Section~\ref{sec:comp}, we calculated properties of the outgoing positron beam assuming that the LXe target is 5.5 radiation lengths thick for a 10 GeV incident electron beam, corresponding to conditions of maximum yield. However, Figure~\ref{fig:Yield} shows that after applying particle selection cuts, the yield is relatively flat for variations in radiation length around the maximum value. On the other hand, Figure~\ref{fig:EDep} illustrates a strong dependence of energy deposited in the LXe target for variations in radiation length near maximum yield. For this reason, we select  shorter radiation lengths that provide nearly-optimal yields while reducing the amount of energy deposited in the target per incident particle. For electron beams with 3 GeV incident energy, the maximum yield after particle selection occurs at 4 radiation lengths, with a yield value of 1.77 and 663 MeV of energy deposited per electron in the target. We select an operating point at 3.5 radiation lengths, which decreases the yield by 6\% while reducing the deposited energy per particle by 24\%. Similarly, the operating point at 4 radiation lengths for 6 GeV corresponds to a decrease in yield of 6\% and reduction in deposited energy of 39\%, and at 10 GeV we select 4.5 radiation lengths which lowers the positron yield by 5\% and reduces the energy deposition by 36\%. We note that target thicknesses in the range of 4 to 4.5 radiation lengths were selected for the NLC and ILC based on similar considerations~\cite{Tang1995,Nagoshi2020}.

After selecting the radiation length operating point for a given incident beam energy, we can calculate the power deposited by the electron beam in the LXe assuming a given bunch charge and repetition rate. We have performed this calculation using beam parameters corresponding to FACET-II, ILC, and C$^3$~\cite{Bai2021}. CLIC is not included in our calculations because they use a two-phase positron production scheme~\cite{Vivoli2010}. 

Table~\ref{tab:BeamInfo} shows the results of the flow rate calculations for the different facilities. For the ILC case, we select the beam parameters from the most recent electron beam-driven source design at 3 GeV, which assumes a target-to-damping ring yield of 1.2, and 50\% overhead of particles delivered to the damping ring~\cite{Nagoshi2020}. The amount of energy deposited in the LXe target by the ILC bunch train is close to, but below, the 96 J$\cdot \textrm{g}^{-1}$ vaporization threshold. Therefore, the LXe flow rate is set by the volume of the LXe target (11.4 cm$^{-3}$) and the bunch train repetition rate of 5 Hz. 

In the case of C$^3$, there is no existing design of the positron source. We assume that the electron train which drives the source has the same time structure as the train of positron bunches that will eventually be fed into the linac. We also assume a target-to-damping ring yield of 1.2, and 50\% overhead of particles delivered to the damping ring, as in the case of the ILC. The C$^3$ bunch train only deposits 2.49 J$\cdot \textrm{g}^{-1}$ in the LXe target, well below the PEDD threshold. Therefore, the LXe target is refreshed at 3.1 Hz, such that the energy deposited by the bunch trains over that interval approaches the PEDD threshold. We note that the LUX-ZEPLIN Dark Matter Experiment uses nearly 2000 liters of LXe and can process 500 liters of xenon gas per minute, which corresponds to roughly 1 liter per minute of LXe~\cite{Akerib2020}. Our application requires 3.4 liters per minute of LXe.

\subsection{Beryllium Windows for the Target Chamber}

\begin{figure}[htb]
    \includegraphics[width = \linewidth]{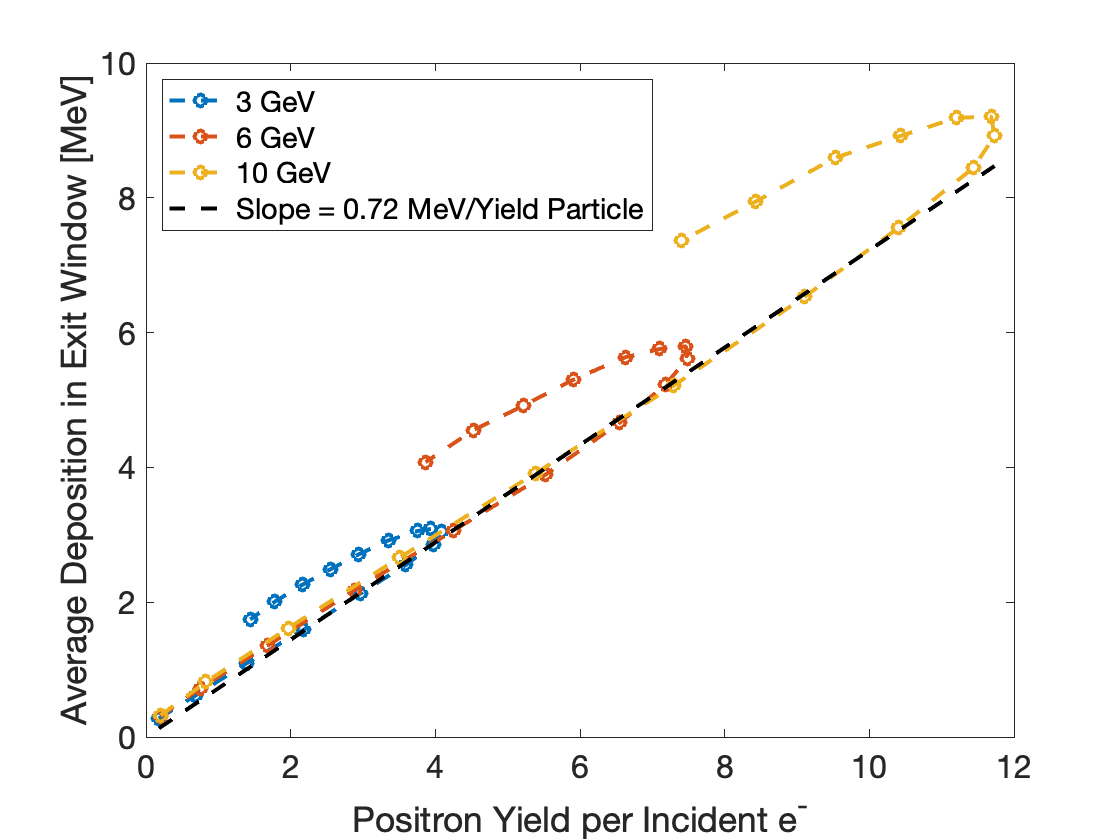}
    \caption{\label{fig:BeDep} Energy deposition in the beryllium exit window as a function of particle yield for three different input electron beam energies. The yield value shown is after the particle selection cuts are applied. There is a linear dependence on energy deposited in the window per yield particle up to the maximum yield for each incoming beam energy.}
\end{figure}
The liquid xenon flows inside a containment vessel which we assume to be made out of steel with entrance and exit windows made out of beryllium (Be). We selected Be as our window material because it is transparent to high energy particles and it has excellent mechanical properties. 

The pressure on the Be windows is dominated by the vapor pressure of LXe which is around 300 kPa. Reference~\cite{Crystran2019} provides a method to calculate the required thickness of the Be windows, which we found to be 0.5 mm using a safety factor of 4, corresponding to 0.001 radiation lengths. Our GEANT4 simulations show that the incoming electron beam deposits a negligible amount of energy in the Be entrance window.

On the other hand, the shower generated by the beam in the LXe target produces a significant flux of low energy particles which interact with the Be exit window. Figure~\ref{fig:BeDep} shows a linear relationship between total target yield and the amount of energy deposited in the exit window per incident particle. For a 3 GeV incident electron beam and 3.5 radiation lengths-thick LXe target, the energy deposited in the Be exit window by the particle shower is 2.86 MeV per incident electron. Since the Be windows are fixed, the total deposited energy is integrated over all particles in the bunch train. For the ILC bunch train, the total energy deposited in the window is 15 J, in excess of the 35 J$\cdot \textrm{g}^{-1}$ PEDD threshold for solid targets. For the C$^3$ bunch train, the PEDD is 1.6 J$\cdot \textrm{g}^{-1}$, well below the safety threshold. From this calculation, we observe that from the perspective of instantaneous energy deposition (energy deposited from a single bunch train), the Be window design is suitable for C$^3$, but the ILC will require a different bunch train format or different window design concept.

In addition to instantaneous energy deposition, we must also take into account the cumulative effect of many bunch trains passing through the material. In particular, can the Be window be cooled fast enough so that it does not reach the melting point of 1285$^{\circ}$ C? The shower power absorbed by the Be window is 75 W for ILC and 57 W for C$^3$.

There are three cooling mechanisms for the Be window: thermal conduction between the window and LXe, thermal conduction between the window and steel vessel, and radiative conduction on the vacuum side of the window. We have calculated the radiative cooling rate to be 0.05 W, which is clearly insufficient for cooling the window. We have yet to calculate cooling rates between the window and LXe, and the window and steel vessel. This will be the subject of future work.

\section{Conclusion}
This work demonstrates that LXe is a viable target alternative for positron sources in HEP applications. Through GEANT4 simulations, we show that a LXe target produces comparable yields to its solid target counterparts. The benefit of using a LXe target is that it does not deteriorate over time as compared to solid targets which require replacement due to degradation. The LXe target has a larger PEDD threshold than solid targets, making it ideal for high power applications, and it is non-toxic which makes it easy to work with compared to liquid metal targets.

The beryllium exit window must be able to tolerate a large amount of absorbed energy from the LXe target shower. This is a key challenge for the LXe target concept which will be further investigated with material and thermodynamic modeling software, such as ANSYS. We will also use ANSYS for computational fluid dynamic modeling of the LXe to understand issues related to cavitation bubbles and turbulent flows.

Future work will couple the output of our GEANT4 simulation to beam capture and transport models, such as GPT and ELEGANT, so that we can fully simulate the target-to-damping ring design.

\section{Acknowledgments}
Work supported by the U.S. Department of Energy under Contract DE-AC02-76SF00515 and the National Science Foundation (Grants No. PHY-1535696 and No. PHY-2012549).

Source code and sample data from GEANT4 simulations can be found at 
\url{https://github.com/MaxVarverakis/LiquidXenonSims.git}.

 \bibliographystyle{elsarticle-num} 
 \bibliography{main}





% \end{thebibliography}
\end{document}